\documentclass{bmcart}

\usepackage{amsthm,amsmath,amssymb}
\RequirePackage[numbers]{natbib}
\RequirePackage{hyperref}
\usepackage[utf8]{inputenc} 

\usepackage{graphicx}

\usepackage{nameref}
\usepackage{url}

\usepackage{array}
\usepackage[inline=true, margin=false]{fixme}

\startlocaldefs
\endlocaldefs

\begin{document}

\begin{frontmatter}

\begin{fmbox}
\dochead{Research}


\title{A Hybrid Adjacency and Time-Based Data Structure for Analysis of Temporal Networks}


\author[
  addressref={aff1},                   
  email={Tanner.Hilsabeck@rockets.utoledo.edu}   
]{\inits{T.H.}\fnm{Tanner} \snm{Hilsabeck}}
\author[
  addressref={aff1},
  email={Makan.Arastuie@rockets.utoledo.edu}
]{\inits{M.A.}\fnm{Makan} \snm{Arastuie}}
\author[
  addressref={aff1},                   
  corref={aff1},                       
  email={Kevin.Xu@utoledo.edu}
]{\inits{K.S.X.}\fnm{Kevin S.} \snm{Xu}}


\address[id=aff1]{
  \orgdiv{Electrical Engineering and Computer Science Department},             
  \orgname{University of Toledo},          
  \city{Toledo},                              
  \state{OH}
  \postcode{43606}
  \cny{USA}                                    
}



\end{fmbox}


\begin{abstractbox}

\begin{abstract} 
Dynamic or temporal networks enable representation of time-varying edges between nodes. 
Conventional adjacency-based data structures used for storing networks such as adjacency lists were designed without incorporating time and can thus quickly retrieve all edges between two sets of nodes (a \emph{node-based slice}) but cannot quickly retrieve all edges that occur within a given time interval (a \emph{time-based slice}). 
We propose a hybrid data structure for storing temporal networks that stores edges in both an adjacency dictionary, enabling rapid node-based slices, and an interval tree, enabling rapid time-based slices. 
Our hybrid structure also enables \emph{compound slices}, where one needs to slice both over nodes and time, either by slicing first over nodes or slicing first over time. 
We further propose an approach for predictive compound slicing, which attempts to predict whether a node-based or time-based compound slice is more efficient. 
We evaluate our hybrid data structure on many real temporal network data sets and find that they achieve much faster slice times than existing data structures with only a modest increase in creation time and memory usage.
\end{abstract}




\end{abstractbox}
%

\end{frontmatter}



\section*{Introduction}
Relational data is often modeled as a network, with nodes representing objects or entities and edges representing relationships between them.
\emph{Dynamic} or \emph{temporal networks} allow nodes and edges to vary over time as opposed to a static network. 
Temporal networks have been the focus of many research efforts in recent years \citep{holme2012temporal, holme2013temporal, holme2019temporal, lambiotte2016guide}. 
Many advancements have been made in temporal network analysis, including development of centrality metrics \citep{nicosia2013graph},  identification of temporal motifs \citep{paranjape2017motifs}, and generative models \citep{junuthula2019block, arastuie2020chip}.

While research on the analysis of temporal networks has advanced greatly, the data structures have seemingly lagged behind. 
A common approach to storing temporal networks is to adopt an adjacency-based data structure for a static network, such as an adjacency list or dictionary, and save timestamps of edges as an attribute, e.g.~using the NetworkX Python package \citep{hagberg2008exploring, hagberg2013networkx}. 

\begin{figure}[t]
    \centering
    \includegraphics[width=4in]{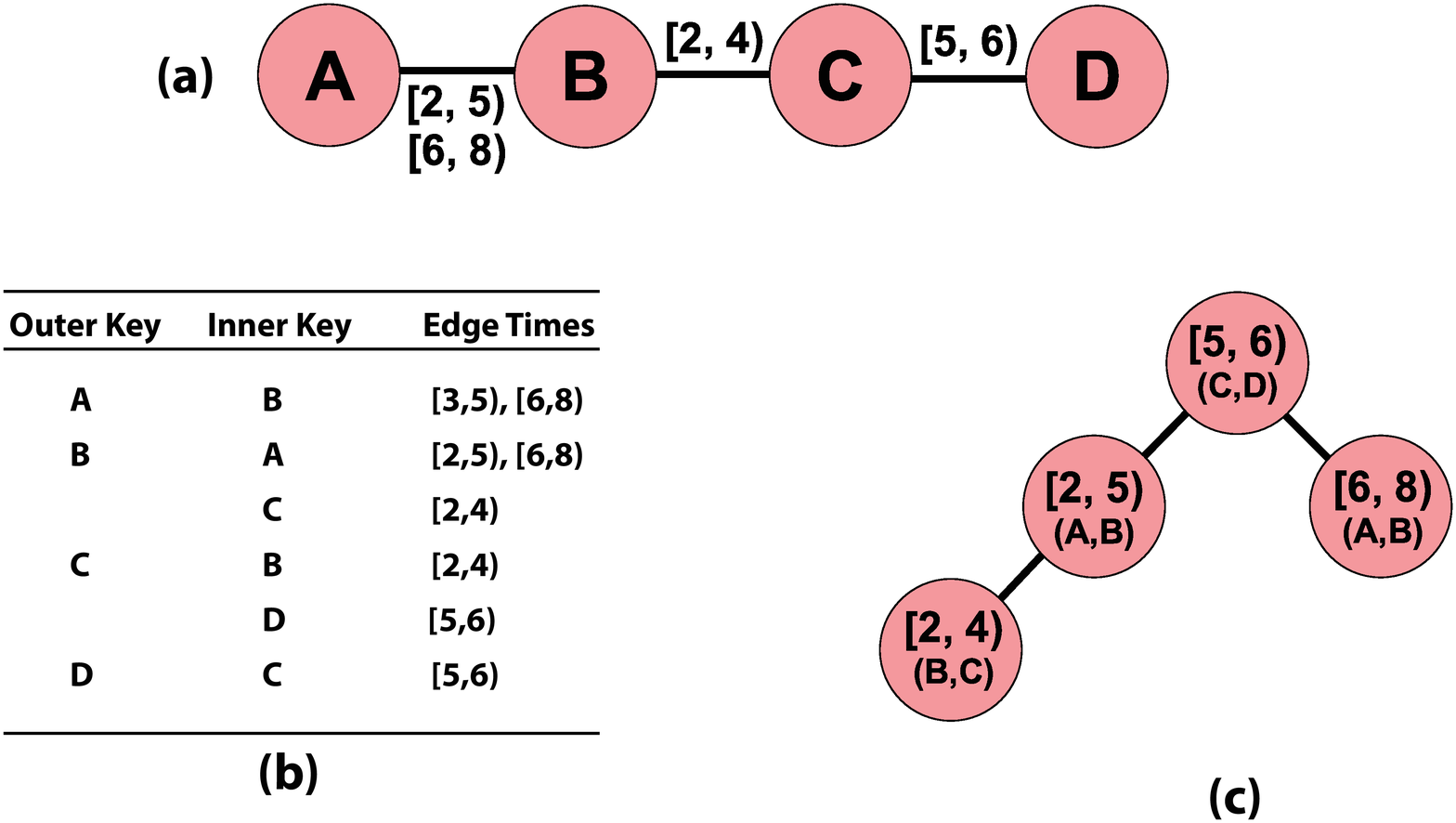}
    \caption{(a) Sample temporal network with 4 nodes and 3 edges. Labels on edges denote time intervals at which the edges are active.
    Our proposed hybrid data structure represents this network using (b) an adjacency dictionary and (c) an interval tree.}
    \label{fig:hybrid_structure}
\end{figure}

In this paper, we design an efficient \emph{hybrid data structure} for temporal networks. 
Our hybrid data structures stores a temporal network using two different data structures: an adjacency dictionary and an interval tree. 
An illustration of our proposed hybrid data structure on a small temporal network is shown in Figure \ref{fig:hybrid_structure}. 
Using two different data structures to store the network has the disadvantage of increased memory usage and increased time to modify a network, i.e.~by inserting or removing edges, which also increases the time to create a network from scratch. 
However, it can significantly decrease the time required to retrieve a subset of the edges in the network, which we refer to as a \emph{slice} of the temporal network. 
In many network analysis tasks, retrieving edges is the main operation dominating computation time \citep{thankachan2018performance}, thus minimizing the slice time is important for analyzing temporal network.

The main objective of our hybrid data structure is to enable rapid slices of three types:
\begin{itemize}
\item \emph{Node-based slices}: Given two node sets $S$ and $T$, return all edges at any time between nodes in $S$ and nodes in $T$. 
Node sets may range from a single node to all nodes in the network. 
Retrieving all temporal edges that contain node $u$ is an example of a node-based slice.
\item \emph{Time-based slices}: Given a time interval $[t_1, t_2)$ or time instant $t$, return all edges between any two nodes that occur in $[t_1, t_2)$ or at time $t$, respectively. Creating a snapshot of a network containing all edges occurring in time interval $[t_1, t_2)$ is an example of a time-based slice.
\item \emph{Compound slices}: Given two node sets $S$ and $T$ as well as a time interval $[t_1, t_2)$ or time instant $t$, return all edges that meet both criteria. 
This can be done by first conducting a node-based or a time-based slice and then retaining only edges that meet both criteria. Creating a partial snapshot of a network containing only a subset of nodes is an example of a compound slice.
\end{itemize}
While adjacency-based structures are excellent for node-based slices, they must iterate over all pairs of nodes to perform a time-based slice. 
This illustrates a fundamental conflict between node- and time-based slices for data structures. 
That is, choosing a data structure that enables rapid node-based slices, such as an adjacency dictionary, results in slow time-based slices, while choosing one that enables rapid time-based slices, such as a binary search tree, results in slow node-based slices. 
This is a limitation of using a single data structure.

Our main contributions in this paper are as follows:
\begin{itemize}
\item We propose a hybrid data structure that stores temporal networks using both an adjacency dictionary
and an interval tree, which enables both rapid node-based and time-based slices.

\item We develop a predictive approach for optimizing compound slices by predicting whether first conducting a node- or time-based slice would be faster given some basic network properties.

\item We demonstrate that our proposed hybrid data structure achieves much faster slice times than existing structures on a variety of temporal network data sets with only a modest increase in creation time and memory usage.
\end{itemize}

The main novelty of our approach is its hybrid structure. 
By using \emph{two separate substructures} (adjacency dictionary and interval tree), we create an approach that is efficiently able to slice across either nodes or time, or both. 
The use of substructures introduces a new problem of ordering when requesting a slice across both nodes and time. 
Our approach also includes a component to \emph{predict} which substructure is more efficient to slice first for a task.

Preliminary results from this paper were presented in the conference publication \cite{hilsabeck2021hybrid}. 
This paper significantly extends those preliminary results in the following ways. 
The main novel contribution in this paper is the development of a new dual linear regression approach for predictive compound slicing along with a thorough investigation of its benefits and drawbacks. 
We also consider a wider range of features to use in the predictive compound slices compared to \cite{hilsabeck2021hybrid}.
We further quantify trade-offs in the prediction accuracy, compound slice times, and creation times as we vary the size of the training data for our predictive model, whereas we focused only on the compound slice times in \cite{hilsabeck2021hybrid}. 
Due to these novel contributions in predictive compound slicing, we have restructured this paper to have a separate section on predictive compound slicing with its own set of experiments. 

\section*{Background and Related Work}
We begin with background information on ways for representing temporal networks and data structures for storing (static) networks. 
We then discuss work related to this paper, including the use of hybrid data structures and other types of temporal network data structures.

\subsection*{Temporal Network Representations}
\label{sec:temporalRepresentations}
Temporal networks are typically represented in one of 3 ways \citep{holme2012temporal, cazabet2020data}:
\begin{itemize}
    \item \emph{Snapshot graph}: a sequence of static graphs, in which an edge exists between nodes $u$ and $v$ if there is an edge active during the time interval $[t_1, t_2)$.
    \item \emph{Interval graph}\footnote{Not to be confused with the other use of the term ``interval graph'' as a graph constructed from overlapping intervals on $\mathbb{R}$ \citep{fulkerson1965incidence}.}: a sequence of tuples $(u, v, t_1, t_2)$ denoting edges between node $u$ and node $v$ during the time interval $[t_1, t_2)$.
    \item \emph{Impulse graph}: a sequence of tuples $(u, v, t)$ denoting edges between node $u$ and node $v$ at the instantaneous time $t$. 
    This representation is also called a contact sequence \citep{holme2012temporal} or a link stream \citep{latapy2017, cazabet2020data}.
\end{itemize}
Snapshot graphs are useful for their ability to quickly restore access to all available static network analysis techniques within each snapshot.
Snapshot graphs typically use \emph{fixed length} snapshots, where 
snapshots 
are taken at regular time intervals (e.g.~every hour) so that finer-grained temporal information is lost within snapshots. 
Thus, snapshot graphs are usually used as a lossy temporal network representation.
However, techniques have been developed in order to minimize this loss by selecting an appropriate number of snapshots based on fidelity and stability metrics \cite{chiappori2021quantitative, leo2019non}.

On the other hand, we consider a \emph{varying length} snapshot representation, where the length of each snapshot may change throughout the duration of the network. 
This differs from the previously mentioned techniques that vary fixed length snapshot intervals in order to obtain desired properties. 
We define the varying length snapshot representation by the creation of a new snapshot upon \emph{each change} in the temporal network.
This allows for a snapshot graph to be a lossless temporal network representation, at the cost of increased complexity and a potentially large number of snapshots, if there are many changes to the network structure over time.
Being lossless is a notable distinction from other representations that vary snapshot intervals over time, such as the aggregation method proposed by Soundarajan et al.~\cite{soundarajan2016} that gradually increases snapshot lengths until the value of a desired graph statistic converges. 
The varying length snapshot representation we consider in this paper does not aggregate over multiple changes to the network.

\subsection*{Data Structures for Networks}
The two main structures for storing a static graph are the adjacency matrix and the adjacency list. 
For a network of $n$ nodes, an adjacency matrix requires $O(n^2)$ space complexity and is thus generally used only for small networks. 
Adjacency lists are typically used instead in many network analysis libraries such as SNAP \citep{leskovec2016snap}.
Adjacency lists can be further improved in average time complexity of most operations (at the cost of a constant factor increase in memory) by using hash tables rather than lists. 
This is sometimes called an \emph{adjacency dictionary} or adjacency map and is the standard data structure in the popular Python package NetworkX \citep{hagberg2008exploring, hagberg2013networkx}.

Static graph structures can be used to store temporal networks by saving time information in edge attributes.
Such structures prioritizes retrieving edges via node-based slices. 
For example, retrieving all edges between nodes $u$ and $v$ can be done in $O(k)$ time using an adjacency dictionary, where $k$ denotes the number of such edges. 
Adjacency-based structures require iterating over all pairs of nodes to conduct a time-based slice, which is slow. 
For example, to retrieve all edges in the time interval $[t_1, t_2)$, one would have to run a double loop over all nodes $u$ and $v$ and retrieve all edges between $u$ and $v$ in time interval $[t_1, t_2)$.

\subsection*{Related Work}

\subsubsection*{Hybrid Data Structures}

Hybrid data structures, which combine different kinds of data structures into one, have a long history in the data structures literature for tasks including searching and sorting \citep{overmars1987design, dietz1982maintaining, korda1999experimental}. 
Such hybrid structures have also recently been proposed for graph data structures, including the use of separate read- and write-optimized structures \citep{thankachan2018performance} and a compile-time optimization framework that benchmarks a variety of data structures on a portion of a data set before choosing one \citep{schiller2015efficient}.

Another related study, although not involving graph data structures, attempts to predict the optimal data structure for a task to minimize its energy consumption \citep{michanan2017greenc5}. 
It generates the prediction by performing energy profiling and training a neural network on features extracted from the energy profiles. 
In this sense, it is similar to our machine learning-based approach for compound slices, which we discuss in the section \nameref{sec:predictive_slices}, although our emphasis is on minimizing computation time, not energy usage.

\subsubsection*{Temporal Network Data Structures}
\label{sec:dynamicstructs}
Most prior work on temporal network data structures has focused on the streaming setting, where the main objective is to design data structures to enable rapid updates to graphs as edges arrive over time in a high-performance computing setting where millions of edges may be changing per second \citep{ediger2012stinger}. 
These types of data structures for massive streaming networks are typically optimized for rapid edge insertions. 
Their objectives differ significantly to those of ``off-line'' analysis of dynamic network data that we consider, where a key objective is to rapidly slice the history of the graph, e.g.~what edges were present at a specific time. 
Indeed it has been found that such high-performance streaming graph structures may be even worse than simple baselines such as adjacency dictionaries for common network analysis tasks including community detection \citep{thankachan2018performance}.

While the focus of this paper is on time-efficient data structures for temporal networks, there has also been prior work on space-efficient structures. 
A fourth-order tensor model proposed by Wehmuth et al.~\citep{wehmuth2014}, which can be expressed by an equivalent square matrix with an index for each time event and elements consisting of a traditional adjacency matrix, is capable of storing dynamic graphs with a memory complexity that scales linearly with the number of edges in a network.
Cazabet \citep{cazabet2020data} considers encoding temporal networks for data compression using the three temporal network representations discussed earlier in the section \nameref{sec:temporalRepresentations}.
We note that it is possible to use both time- and space-efficient structures as part of a complete workflow by storing data using the more space-efficient format, while loading it into memory to be analyzed using the more time-efficient format, but it is beyond the scope of this paper.

\subsection*{Temporal Networks Software Packages}
There are multiple Python software packages that support temporal networks.
DyNetX \cite{giulio_rossetti_2021_5599265} and tnetwork \cite{cazabet_2022_03_21}, two packages built on NetworkX \citep{hagberg2008exploring, hagberg2013networkx}, provide users access to commonly used methods when working with temporal networks that are not featured in the generic graph classes.
tnetwork contains multiple classes capable of storing temporal networks.
While providing a standard class for snapshot representation, the impulse and interval representations store temporal information as edge attributes in a NetworkX graph.
DyNetX stores temporal information in a series of snapshot graphs.
However, unlike a traditional snapshot method, edges are not saved in each snapshot in which they appear, but only the snapshots in which they begin and end.
This approach significantly reduces the amount of memory required to store the network, but requires a search through all temporal edges within a network to perform a temporal slice.

Teneto \cite{william_hedley_thompson_2020_3626827} and Tacoma \cite{maier2020} are two more examples of Python packages aimed at temporal network analysis.
Both of these packages store their network as unsorted edge lists.
Therefore, they require time linear in the total number of temporal edges in the network to perform a temporal slice, regardless of the size of the slice.

pathpy \cite{scholtes2017network} is the most similar currently available Python package to our proposed structure due to its implementation of another Python package named intervaltree \cite{halbert2020}.
However, despite its shared name, intervaltree's structure is not equivalent to our interval tree implementation.
According to the documentation, intervaltree performs a temporal slice with a time complexity of $O(r \log m)$, where $r$ is the number of timestamps between the requested start and end time.
It achieves this by performing multiple single point slices for every overlapping timestamp.
This approach results in multiple traversals of the tree for to slice a time interval.
The intervaltree package also restricts the duration of intervals, specifying that their duration must be greater than zero.
Therefore, impulses are not allowed in intervaltree or pathpy.

\begin{table}[tp]
\centering
\caption{Comparison of Python packages for temporal networks. Check marks represent package's ability to store a temporal network using the corresponding representation. Interval implementation gives a brief description of the data structure used to store interval graphs.}
\label{tab:package_comparison}
\begin{tabular}{ccccm{1.9in}}
\hline
Package & Snapshot & Impulse & Interval & Interval graph structure \\
\hline
DyNetX \cite{giulio_rossetti_2021_5599265} & \checkmark &  &  & None \\
tnetwork \cite{cazabet_2022_03_21} & \checkmark & \checkmark & \checkmark & Adjacency dictionary with timestamps as edge attributes \\
Teneto \cite{william_hedley_thompson_2020_3626827} & & \checkmark &  & None \\
Tacoma \cite{maier2020} & & \checkmark & \checkmark & Edge list with timestamps \\
pathpy \cite{scholtes2017network} & & & \checkmark & Interval tree \cite{halbert2020} \\
DyNetworkX \cite{hilsabeck20} & \checkmark & \checkmark & \checkmark & Hybrid: adjacency dictionary + interval tree \\
\hline
\end{tabular}

\end{table}

Our proposed hybrid data structure is implemented in IntervalGraph class in the Python package DyNetworkX \cite{hilsabeck20}. 
Similar to tnetwork, DyNetworkX also has classes available to store snapshot and impulse graphs. 
A comparison of the different software packages and their data structures used is provided in Table \ref{tab:package_comparison}.

\section*{Proposed Hybrid Data Structure}
Our proposed data structure to store temporal networks is a hybrid data structure consisting of an adjacency dictionary and interval tree, as shown in Figure \ref{fig:hybrid_structure}.
Our main objective in designing the hybrid data structures for temporal networks is to rapidly retrieve a subset of edges that meet specified criteria, which we call \emph{slicing}. In the off-line analysis setting that we target, retrieving edges is the main operation dominating computation time of network analysis tasks \citep{thankachan2018performance}, so rapid slices are more important than rapid insertions or deletions.

The advantage of our proposed hybrid data structure is that one can choose either the adjacency dictionary or the interval tree, whichever one is faster, to perform the slice. 
The main disadvantage is that any changes to the network, i.e.~insertion or removal of edges requires modifying two data structures, which is slower than modifying only one data structure. 
Another disadvantage is increased memory usage from storing two separate data structures. 
It should be mentioned that, although we utilize two data structures, memory location pointers are used to avoid duplicating the data; therefore, memory usage is only modestly increased to store the data structure itself.

\subsection*{Interval Tree: Time-based Slices}
The first novel component of our hybrid data structure is an interval tree to store edges using the edge time duration $[t_1, t_2)$ as the key. 
For instantaneous edges at time $t$, we use the trivial interval $[t,t]$. 
Interval trees can be implemented as an extension of a variety of self-balancing binary search trees, including red-black trees and AVL trees \citep{cormen2009introduction}.
For our purpose, we select the AVL tree as the base representation of our interval tree in hopes of maximizing performance during slices due to its more rigid balancing algorithm.
The size of the interval tree is equal to the number of unique intervals and impulses in a data set.

We use the interval tree structure to perform \emph{time-based slices}, which retrieve all edges \emph{between any two nodes} with times $[t_1, t_2)$ that overlap a given search interval $[s_1, s_2)$. 
A time-based slice for an instantaneous time $s$ can also be performed using an interval $[s, s]$. 
Once the tree is traversed, each edge time determined to be overlapping with the search interval yields all edges stored within. 
Given a temporal network with $m$ edges, the interval tree has space complexity $O(m)$ and search time complexity of $O(\log m + k)$, where $k$ denotes the number of edges that meet the search criteria \citep{lee2005interval}.

An example traversal for a time-based slice over the interval [3, 4) on the interval tree in Figure \ref{fig:hybrid_structure}(c) is as follows:
\begin{enumerate}
\item First, the root [5, 6) is evaluated to see if it overlaps the requested interval, it does not.
\item Then, we check the children, [2, 5) and [6, 8). [2, 5) overlaps the requested interval [3, 4), therefore, it is returned.
[6, 8) does not overlap [3, 4) and traversal terminates along that path.
\item Finally, the grandchild of the root is checked, [2, 4), which is returned as well.
\end{enumerate}

\subsection*{Adjacency Dictionary: Node-based Slices}
The second part of our hybrid structure is an adjacency dictionary, an adjacency list-like structure implemented using nested hash tables rather than lists, similar to the NetworkX Python package \citep{hagberg2008exploring, hagberg2013networkx}.
The outer table stores the keys associated with the an edge's first node, and the inner table stores keys representing an edge's second node.
The inner table's values hold a list of all edge times containing the corresponding node pair. 
For directed networks with edges from $u$ to $v$, two separate nested hash tables are created: the first with outer keys $u$ and inner keys $v$, the second with outer keys $v$ and inner keys $u$.

We use the adjacency dictionary to perform \emph{node-based slices}, which retrieve all edges \emph{at any time} between two node sets $S$ and $T$. 
Either of the sets could range from a single node to the set of all nodes. 
For example, if $S$ denotes a single node while $T$ denotes the set of all nodes, then the node-based slice is enumerating all edge times with neighboring nodes of $S$. 
Since the nested dictionary contains a list of all edge times, the space complexity of this structure is $O(m)$, and the search time complexity is $O(k)$.

\subsection*{Compound Slices}
\label{sec:compound}
A \emph{compound slice} retrieves all edges between two node sets $S$ and $T$ with times $[t_1, t_2)$ that overlap a given search interval $[s_1, s_2)$. 
(Either the edge times or search interval could also be instantaneous as with time-based slices.)
It combines both the criteria of the time-based and node-based slices. 

A compound slice can be performed in two ways. 
The first is to perform a node-based slice using the adjacency dictionary, returning all edges between node sets $S,T$, and then filter the edges based on the search interval. 
The second is to perform a time-based slice using the interval tree, returning all edges overlapping the search interval $[s_1, s_2)$ between any two nodes, and then filter the edges based on the node sets. 
Depending on the node sets and search interval, one approach for compound slicing may be faster than the other. 
We defer discussion on how to predict the faster approach for compound slicing to the section \nameref{sec:predictive_slices}.

\subsection*{Experiments}
We are primarily interested in the off-line analysis setting where an entire network is first loaded into memory and then different analysis tasks are performed by slicing the data structure. 
To evaluate the effectiveness of our proposed hybrid data structure, which we refer to as IntervalGraph, we are primarily interested in the time required to compute a time-based slice. 
Since our adjacency dictionary structure is almost identical to a typical adjacency dictionary, e.g.~in NetworkX \citep{hagberg2008exploring, hagberg2013networkx}, we do not evaluate node-based slices. 
Of secondary interest are the creation (load) time from a text file and memory usage, both which we expect to be slightly higher than the comparison baselines due to maintaining two data structures. 
We perform experiments measuring these quantities on six real data sets on our proposed IntervalGraph hybrid data structure and compare them against several other data structures.
Unless otherwise specified, each structure and data set combination is recorded and averaged 100 times in order to reduce variance in CPU clock rate between measurements\footnote{All experiments were run on a workstation with 2 Intel Xeon 2.3 GHz CPUs, totaling 36 cores, and 384 GB of RAM on Python version 3.8.3.}. 

\subsubsection*{Data Sets}

\begin{table}[tp]
\centering
\caption{Data sets used for evaluation. 
Edges denote temporal edges, which count edges between a pair of nodes at multiple times as multiple edges. 
Edge durations shown are the mean over all pairs of nodes with at least one edge.}
\label{tab:interval_data}
\begin{tabular}{cccccc}
\hline
Data set        & Nodes   &  Edges     & Resolution & Directed? & Edge Duration \\
\hline
Enron \citep{priebe05,priebe09}  &   184 &   125,235 &   1 second    &   Yes &   0 (Impulses) \\
Bike share \cite{cycling21} & 793 &   9,882,954   &   1 minute    &   Yes &   21.1 minutes\\
Reality Mining \citep{eagle2006reality, eagle2009inferring}  & 6,809   & 52,050    & 1 second   & Yes       & 176.1 seconds \\
Infectious \cite{isella2011s}     & 10,972  & 198,198   & 20 seconds & No        & 41.97 seconds \\
Wikipedia links \cite{ligtenberg2017introduction} & 43,509  & 160,797    & 1 second   & Yes       & 2.63 years \\
Facebook wall \cite{viswanath2009evolution}  & 43,953  & 852,833    & 1 second   & Yes       & 0 (Impulses) \\
Ask Ubuntu  \cite{leskovec2014snap, paranjape2017motifs}    & 159,316 & 964,437  & 1 second   & Yes       & 0 (Impulses) \\
\hline
\end{tabular}

\end{table}

We evaluate the proposed models using the real temporal network data sets shown in Table \ref{tab:interval_data}. 
Additional details on these data sets are provided in the following.
\begin{itemize}
\item \emph{Enron \citep{priebe05,priebe09}}: This data set contains timestamped emails from the Enron email corpus. 
We consider all emails either sent to, cc, or bcc a recipient as edges.
\item \emph{Bike share \cite{cycling21}}: This data set contains the usage of publicly available bikes available for rent in and around the city of London, England.
Data is collected and maintained by Transport for London.
In this paper we will be limiting this data set to trips occurring in 2016.
\item \emph{Reality Mining \citep{eagle2006reality, eagle2009inferring}}: The data set was collected by logging the smartphone activities of 94 test subjects at MIT. 
We use the voice calls data, where the events are initiated or received calls with durations. 
Users who were not part of the test subjects but received calls from at least one of the test subjects are also included.
\item \emph{Infectious \cite{isella2011s}}: The data set contains face-to-face proximities between visitors to a museum exhibit collected using the active RFID-based SocioPatterns platform \citep{cattuto2010dynamics}, which scans every 20 seconds. 
\item \emph{Wikipedia links \cite{ligtenberg2017introduction}}: The events in the Wikipedia links data set are the times at which links between different Wikipedia pages are added and removed. 
We consider the subset of the data used in \cite{platt2019network} to enable a direct comparison in the section \nameref{sec:exp_casestudy}.
\item \emph{Facebook wall \cite{viswanath2009evolution}}: The data set consists of timestamps of wall posts between Facebook users in the New Orleans region from September 2004 to January 2009. 
We consider the largest connected component, which contains about 94\% of the nodes and 97\% of the wall posts. 
We also consider only posts from a user to another user's wall so that there are no self-edges.
\item \emph{Ask Ubuntu \cite{leskovec2014snap, paranjape2017motifs}}: The data set contains three types of timestamped interactions on the Stack Exchange web site Ask Ubuntu: a user $u$ answering user $v$'s question, $u$ commenting on $v$'s question, and $u$ commenting on $v$ answer.
\end{itemize}

In the Wikipedia data, events are expected to last a long time (months to years), while events in the Bike share, Reality Mining, and Infectious data are more likely to have short durations on the order of seconds to minutes. 
The Enron, Facebook wall, and Ask Ubuntu data sets contain instantaneous events, while the other data sets contain interval events. 
These data sets span a wide range in terms of size, time resolutions, and duration of events, ranging from networks with very few nodes but lots of short temporal edges (London bike share), to networks with lots of nodes and extremely long duration temporal edges (Wikipedia links).

\subsubsection*{Comparison Baselines}
Four other data structures will serve as our baselines for comparison with our proposed hybrid structure.
The first structure is a MultiGraph in NetworkX \citep{hagberg2008exploring, hagberg2013networkx}, with intervals stored as edge attributes, representing the de facto standard for network structures in Python.
This structure is representative of performance using only an adjacency dictionary. 
The second structure, SnapshotGraph, is the variable window snapshot technique described in the section \nameref{sec:temporalRepresentations}. 
Snapshots are stored in a SortedDictionary from Python package Sorted Containers \citep{jenks2019python}.
The third structure, AdjTree, is an adjacency dictionary with internal elements consisting of an interval tree for each node pair.
This baseline represents a simplified single structure approach (rather than the hybrid IntervalGraph structure that we propose). 
The last baseline, TVG, is the fourth-order tensor model by Wehmuth et al.~\citep{wehmuth2014} described in the section \nameref{sec:dynamicstructs}.
In order to assist with slicing, the matrix representation has been adapted into dictionary equivalents.
As implemented, the structure consists of a SortedDictionary storing $t_1$ keys, with values pointing to SortedDictionaries containing $t_2$.
The second dictionary points to a standard adjacency dictionary.

\subsubsection*{Time-Based Slices}
\label{sec:time_slice}

\begin{figure}[tp]
\centering
\includegraphics[width=4.8in]{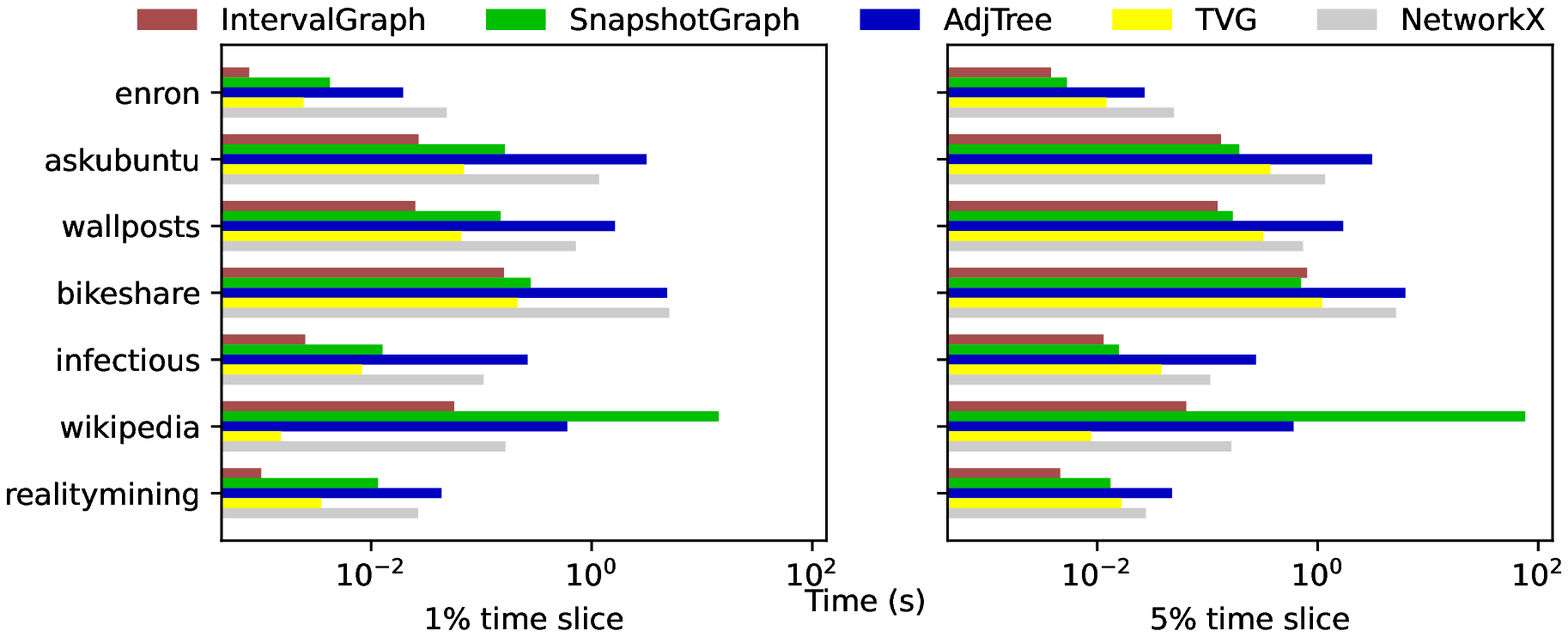}
\caption[Time-based slice times for 1\% and 5\% time slices across all data sets]{Time-based slice times for 1\% and 5\% time slices across all data sets.
The proposed hybrid interval tree structure has significantly lower slice times on most data sets.}
\label{fig:time_slices}
\end{figure}

Figure \ref{fig:time_slices} compares the time to return all edges within 1\% and 5\% time slices of the network duration. 
On such small time slices, especially at 1\%, our proposed interval tree-based structure, IntervalGraph, is far superior to the other structures on almost all of the data sets. 
The exception is for the Wikipedia data, where TVG is the superior structure. 
The Wikipedia data set is quite different from the other data sets in that the mean edge duration is about 2.6 years or 27\% of the length of the total data trace. 
This is extremely large compared to the Reality Mining data, which is a more typical data set, where the mean edge duration is about 3 minutes or 0.002\% of the trace length. 
SnapshotGraph performs extremely poorly on the Wikipedia data, with slice times exceeding even the NetworkX MultiGraph baseline, while AdjTree is a poor performer on all of the data sets.

Compared to the NetworkX MultiGraph baseline, the improvement in slice time offered by IntervalGraph is typically over one order of magnitude for both the 1\% and 5\% slices. 
Here, we see the clear benefit of the hybrid structure allowing us to use the more beneficial data structure, the interval tree, to perform the time-based slice.

\subsubsection*{Creation Time and Memory Usage}

\begin{figure}[tp]
\centering
\includegraphics[width=4.8in]{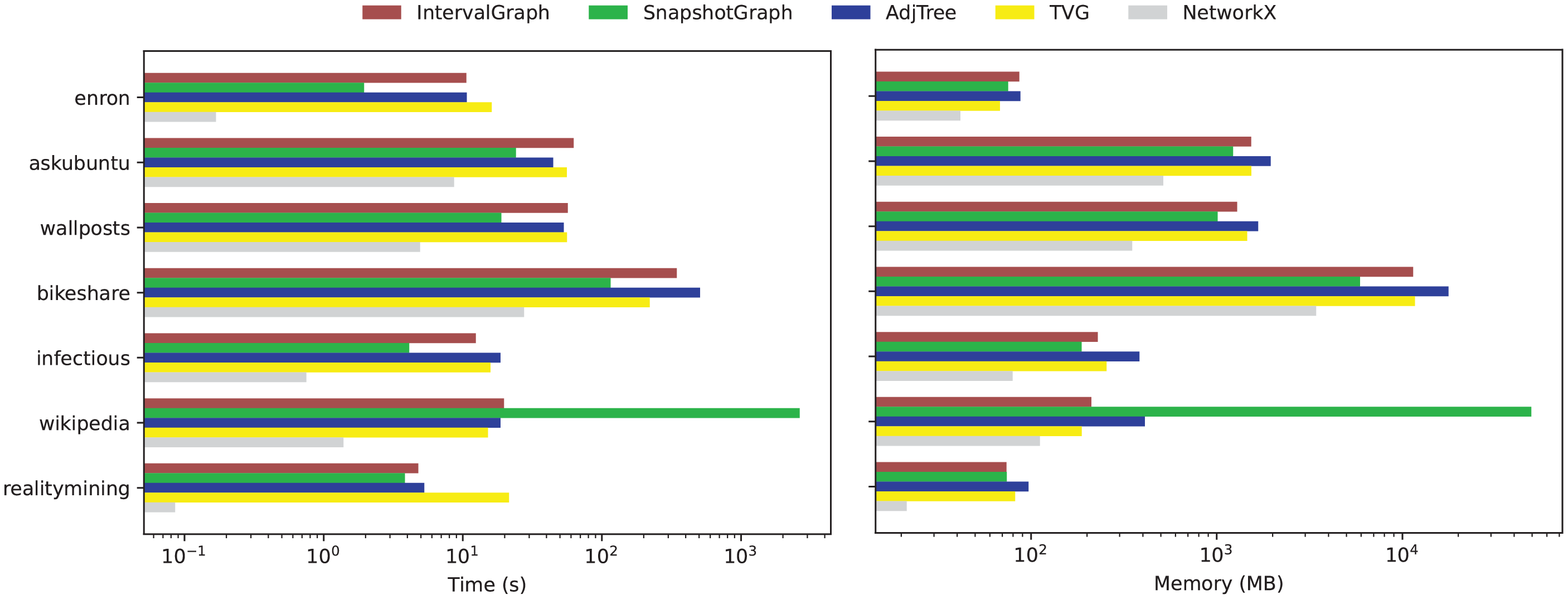}
\caption[Creation times and memory usage for different temporal network data structures]
{Creation times and memory usage for different temporal network data structures.
The NetworkX MultiGraph is the simplest data structure and has the lowest creation time and memory usage (at the cost of slow time-based slice times as shown in Figure \ref{fig:time_slices}). 
All other data structures have roughly comparable creation times and memory usage.}
\label{fig:creationmem}
\end{figure}

A comparison of creation time and memory usage for the different structures is shown in Figure \ref{fig:creationmem}. 
With its lack of sorted edges with respect to time, NetworkX has a creation time of at least one order of magnitude faster than the second fastest data structure on most data sets, SnapshotGraph.
SnapshotGraph struggles to efficiently store edges that extend across a large number of snapshots, resulting in a memory usage of over 49 GB on the Wikipedia data {set.}
The remaining three structures (IntervalGraph, AdjTree, and TVG) tend to have creation times between $\pm$10\% of each other, depending on the data set. 
This trend continues when examining memory usage of each structure, where these three structures continue to be within $\pm$10\% of each other on most data sets.
However, the difference in memory usage between NetworkX and these three structures shrinks to a factor of 2-3x.

The increased creation time and memory usage for IntervalGraph compared to the NetworkX MultiGraph baseline illustrates the trade off that accompanies the fast time-based slices that IntervalGraph provides. 
While the slice time is incurred every time one needs to retrieve edges from the temporal network, the creation time is a one-time cost, and the improved slice time rapidly compensates for the increase creation time of IntervalGraph.

\subsubsection*{Case Study}
\label{sec:exp_casestudy}
In an ideal world, a data analyst would spend the majority of his or her time analyzing data.
However, in reality, an increasing large portion of time is spent creating and slicing the data before analysis can even begin.
In this case study, we will evaluate the computation time of a sample data analysis workflow using IntervalGraph and NetworkX on the London Bikeshare data set.
To begin the analysis, a one-time upfront computation cost must be paid in order to create the additional data structures of IntervalGraph and NetworkX.

Analysts often wish to determine how network metrics change over time, which requires frequent slicing of the data set.
In this example, we wish to calculate the daily betweenness centrality across all nodes, so 365 slices are required.
Slice time represents the total time required to retrieve all edges for all slices.
Only once the slicing of the data structure occurs can the analysis begin.
While the analysis task performed in this case study is betweenness centrality via the NetworkX package, it should be noted that the exact analysis task has no impact on performance as the slicing process returns an identical list of edges.

\begin{figure}[tp]
\centering
\includegraphics[width=4in]{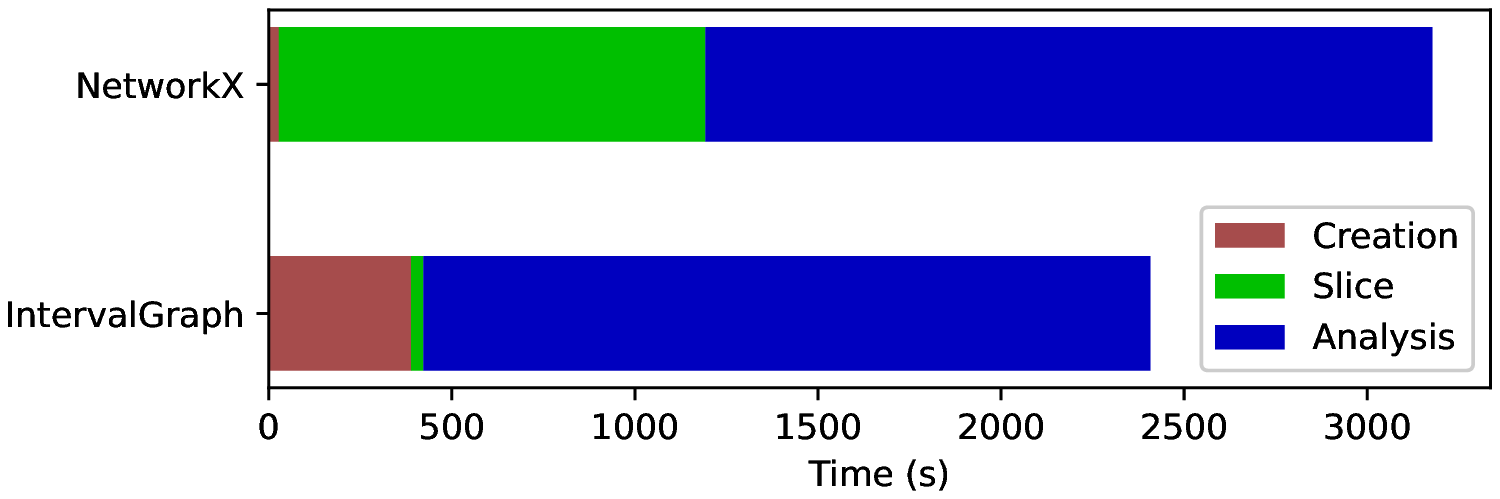}
\caption{Computation time per stage on the London Bikeshare data set. 
The cost of the increased creation time for IntervalGraph rapidly disappears when performing many time-based slices, which are much faster than IntervalGraph.}
\label{fig:casestudy}
\end{figure}

Computation times per stage for our proposed hybrid IntervalGraph structure and NetworkX are located in Figure \ref{fig:casestudy}.
IntervalGraph's creation time is much longer than the NetworkX implementation due to its tree sub-structure.
However, it is important to remember this cost must only be paid once as the object may be loaded from permanent storage once the temporal edges are sorted.
For small analysis tasks requiring little slicing, IntervalGraph's ability to more efficiently retrieve temporal edges may not outweigh this large upfront cost.
With the 365 slices performed in this case study, IntervalGraph is almost 3 times faster than NetworkX after creation and slicing. 
This speed up translates to a 25\% reduction in computation time over the entire workflow, including the analysis time.
Depending on the size of the network, we find that IntervalGraph becomes more efficient at completing the overall workflow in anywhere from 5 to 100 slices.
We believe this number of slices is low enough to make IntervalGraph more efficient than NetworkX in most use cases, especially during the exploratory analysis stage where a wide variety of snapshot lengths may be sliced.

\section*{Predictive Compound Slices}
\label{sec:predictive_slices}
We now consider the task of minimizing computation time for a compound slice to retrieve all edges between two node sets $S$ and $T$ with times $[t_1, t_2)$ that overlap a given search interval $[s_1, s_2)$. 
As discussed in the section \nameref{sec:compound}, there are two ways to perform a compound slice: by first performing a node-based slice on the node sets $S$ and $T$ or by first performing a time-based slice on the search interval $[s_1, s_2)$. 
Whether first performing a node-based slice or time-based slice results in a faster slice time depends on the node sets and the search interval.
Intuitively, one might expect that a compound slice over large node sets and a small search interval (or in the limiting case, a single time instant $s$) would be faster by first performing a time-based slice. 
Similarly, one might expect that a compound slice over small node sets (or in the limiting case, single nodes $u$ and $v$) would be faster by first performing a node-based slice. 
Therefore, when tasked with a compound slice, an ideal hybrid structure should attempt to \emph{predict the correct sub-structure} to use in order to achieve optimal time efficiency. 

We propose to predict whether a node-based slice or time-based slice is faster by running a small set of training slices when loading a data set into memory. 
For these training slices, we run both types of compound slices, node-based and time-based, and record their respective slice times. 
We compute features from these training slices, which we then use to train a machine learning model to predict the compound slice times, which we describe in the following.

\subsection*{Predicting Compound Slice Times}

\subsubsection*{Feature Construction}
\label{sec:features}
For each compound slice, we compute the following features.
The first two features, \texttt{percentOfNodes} and \texttt{percentOfInterval}, correlate to the number of nodes and length of interval, respectively, specified by the slice.
The next feature is \texttt{sumOfDegrees}, representing the number of temporal edges returned by a node-first slice.
Lastly, a \texttt{lifespan} is calculated for each node by normalizing the time between a node's first and last appearance with respect to the network's trace length.

In addition to the network features described above, we also attempt to extrapolate the expected number of edges returned from a subgraph consisting of 1\% of the total temporal edges, selected at random.
The subgraph is temporally sliced into 100 intervals, where each \emph{nth} slice corresponds to the \emph{nth} percent of the original trace.
Three features are calculated from subgraph's temporal slices by finding the minimum, maximum, and mean number of edges returned from all intervals overlapping the initial slice.
We refer to these features as \texttt{binMin}, \texttt{binMax}, and \texttt{binMean}.

The time required to compute the features differs depending on the type of feature, with \texttt{percentOfNodes} and \texttt{percentOfInterval} being the fastest. 
We consider three sets of features for evaluation:
\begin{enumerate}
\item \texttt{percentOfNodes} and \texttt{percentOfInterval}.
\item \texttt{percentOfNodes},  \texttt{percentOfInterval}, \texttt{sumOfDegrees}, and \texttt{lifespan}.
\item \texttt{percentOfNodes},  \texttt{percentOfInterval}, \texttt{sumOfDegrees}, \texttt{lifespan}, \texttt{binMin}, \texttt{binMax}, and \texttt{binMean}.
\end{enumerate}

\subsubsection*{Predictive Models}
We consider two different predictive models for the compound slice times. 
First, we formulate the predictive compound slicing problem as a binary classification problem, where the binary prediction target is whether the node-based slice is faster than the time-based slice. 
Second, we consider a dual linear regression model, where we have two separate linear regression models to predict the compound slice times using the node-based slice first and using the time-based slice first. 

In order to create training data for our predictive models, we create samples via the steps below.
First, we randomly select from 1\% to 50\% of the network's nodes and trace length, respectively.  
We then randomly select the node set and search interval given the previous percentages.
Two compound slices are then performed, one using the node-based slice first, the other using the time-based slice first. 
Compound slices which return zero edges are rejected from the training set and replaced with a new random sample. 
We compute features for each randomly generated slice. 
For the dual linear regression model, the recorded computation times of both the node-based and time-based slices are used as the prediction targets. 
For the logistic regression model, the prediction target is binary, indicating whether the node-based slice is faster than the time-based slice. 

While our proposed structure remains usable via a user-defined order for compound slices without a model, in order to automate the benefit of the hybrid-structure approach, the time required to create the model, must be included.
The time required to generate the training data dominates the overall time required to create the model as a number of inefficient slices must be made. This effect is more pronounced on data sets with a high edge-to-node ratio as the node-based slice returns a large amount of edges. 
We initially use 50 training slices and compute only the 2 fastest features \texttt{percentOfNodes} and \texttt{percentOfInterval}. 
We vary both the number of slices and the number of features in our experiments to assess their effects on both the compound slice and creation times. 
Since only the coefficients and intercepts of the linear regression must be saved in order to predict future slices, the memory increase of the structure due to the predictive component is negligible compared to the memory usage of the data sets from Figure \ref{fig:creationmem}, which is typically on the order of 100 MB to 10 GB.

\subsection*{Experiments}
We are primarily interested in the compound slice time for our predictive compound slicing models. 
We also investigate the accuracy of our predictive models and the effects of including a predictive model on the creation time and memory usage.

\subsubsection*{Compound Slice Time}

\begin{figure}[tp]
\centering
\includegraphics[width=4in]{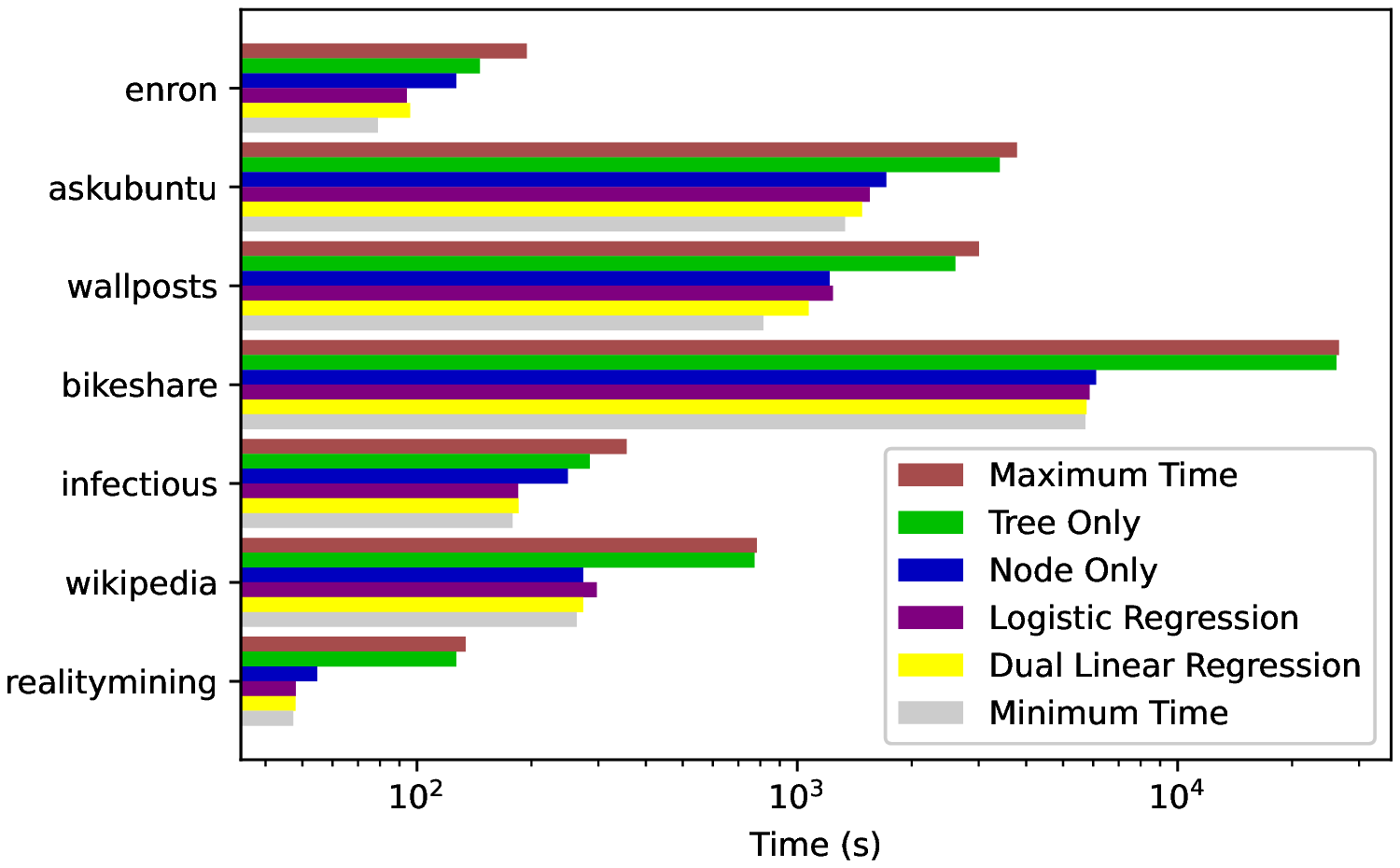}
\caption{Compound slice times for different approaches on IntervalGraph. 
Our proposed logistic and dual linear regression predictive compound slicing approaches perform better than using only node- or tree-based slices.}
\label{fig:compound}
\end{figure}

Recall that there are two ways to perform a compound slice: a node-based compound slice using the adjacency dictionary and a time-based compound slice using the interval tree. 
In Figure \ref{fig:compound}, we compare the compound slice times using four strategies: always using a node-based slice, always using a time-based slice, and using our two prediction models (logistic regression and dual linear regression) of which slice is faster. 
We compare these to the minimum and maximum times, i.e.~always selecting the faster or slower approach respectively, that would be could be achieved (which are not known in practice).

Upon analysis, we find that node-only strategy is faster than the tree-only strategy across all tested data sets. 
However, for some individual slices, time-based slicing is faster, which is why the node-only time is not necessary the minimum time.
Note that our proposed dual linear regression approach is faster than either the node-only or tree-only approaches on all of the data sets.
On many data sets, it approaches the minimum possible time. 
Our logistic regression model is slightly worse than dual linear regression, suggesting that knowing the actual training slice times from each compound slice approach carries additional value compared to just knowing which type of compound slice is faster. 
Thus, we consider only dual linear regression in the following.

\subsubsection*{Prediction Accuracy}

\paragraph{Training Set Size}

\begin{figure}[tp]
\centering
\includegraphics[width=4.8in]{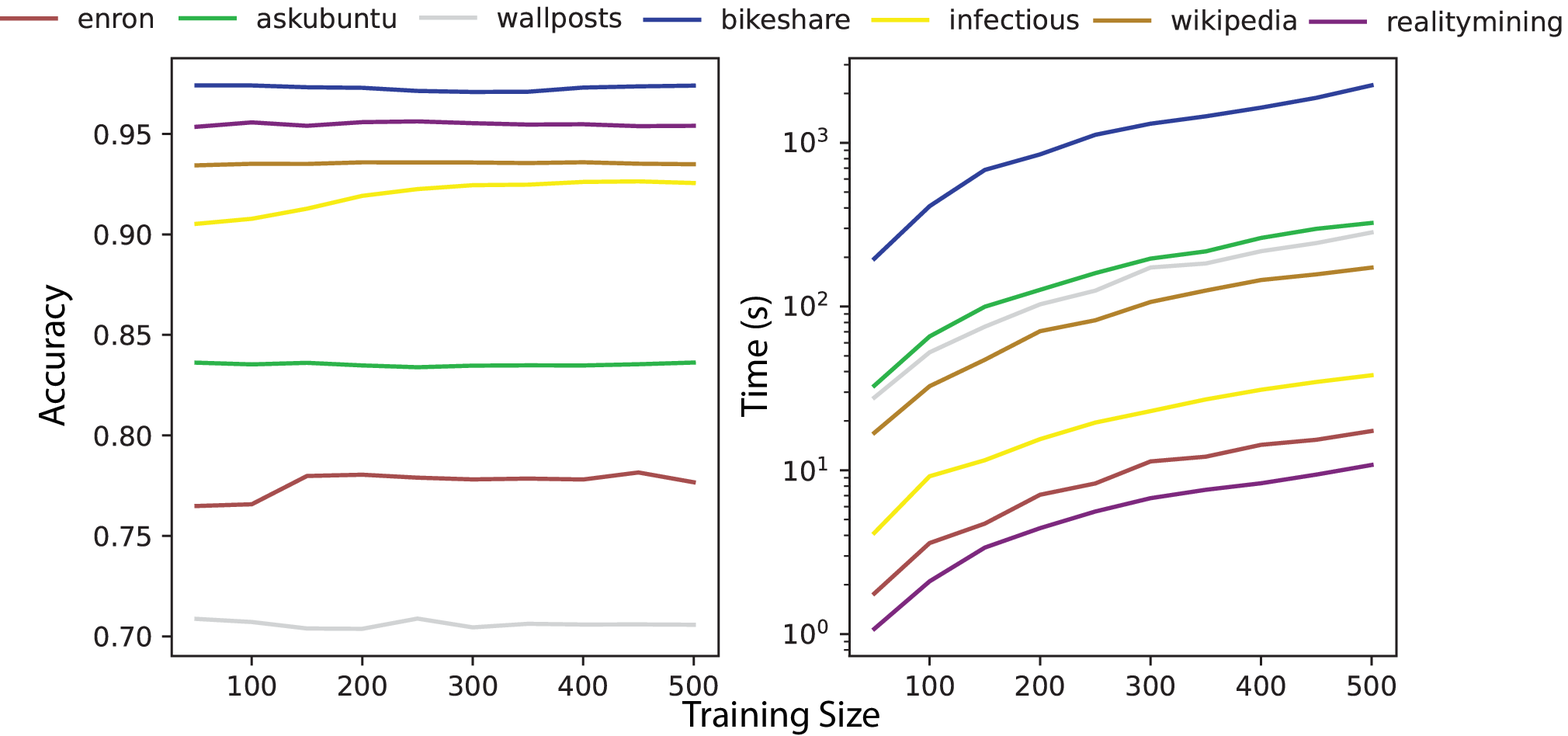}
\caption{Prediction accuracy and computation time for the dual linear regression model as the number of training slices is varied.
On most data sets, the prediction accuracy does not significantly increase as the training size is increased, so using more training slices has minimal benefit.}
\label{fig:accuracy_training_size}
\end{figure}

We next evaluate the prediction accuracy of our dual linear regression model as we vary the size of the training set, i.e.~the number of slices we use to train the model.  
The size of the training set is important because it takes time to compute the slices to form the training set, which is added to the creation time for the hybrid structure. 
Thus, the prediction objective is not necessarily to achieve the maximum prediction accuracy as is typical in machine learning applications.
Increasing the size of the training set to achieve marginal gains in prediction accuracy may not result in appreciable decreases in compound slice times while increasing creation time. 
The prediction accuracy of the dual linear regression model and computation time for the training slices as the training size is varied are shown in Figure \ref{fig:accuracy_training_size}. 
Notice that prediction accuracy does not appear to improve at all on 5 of the 7 data sets, with improvements visible only on the Infectious and Enron data sets. 
Thus, increasing the training size is mostly adding to the creation time without any additional benefit, and we do not see a need to use more than 50 training slices.

\begin{figure}[tp]
\centering
\includegraphics[width=4.8in]{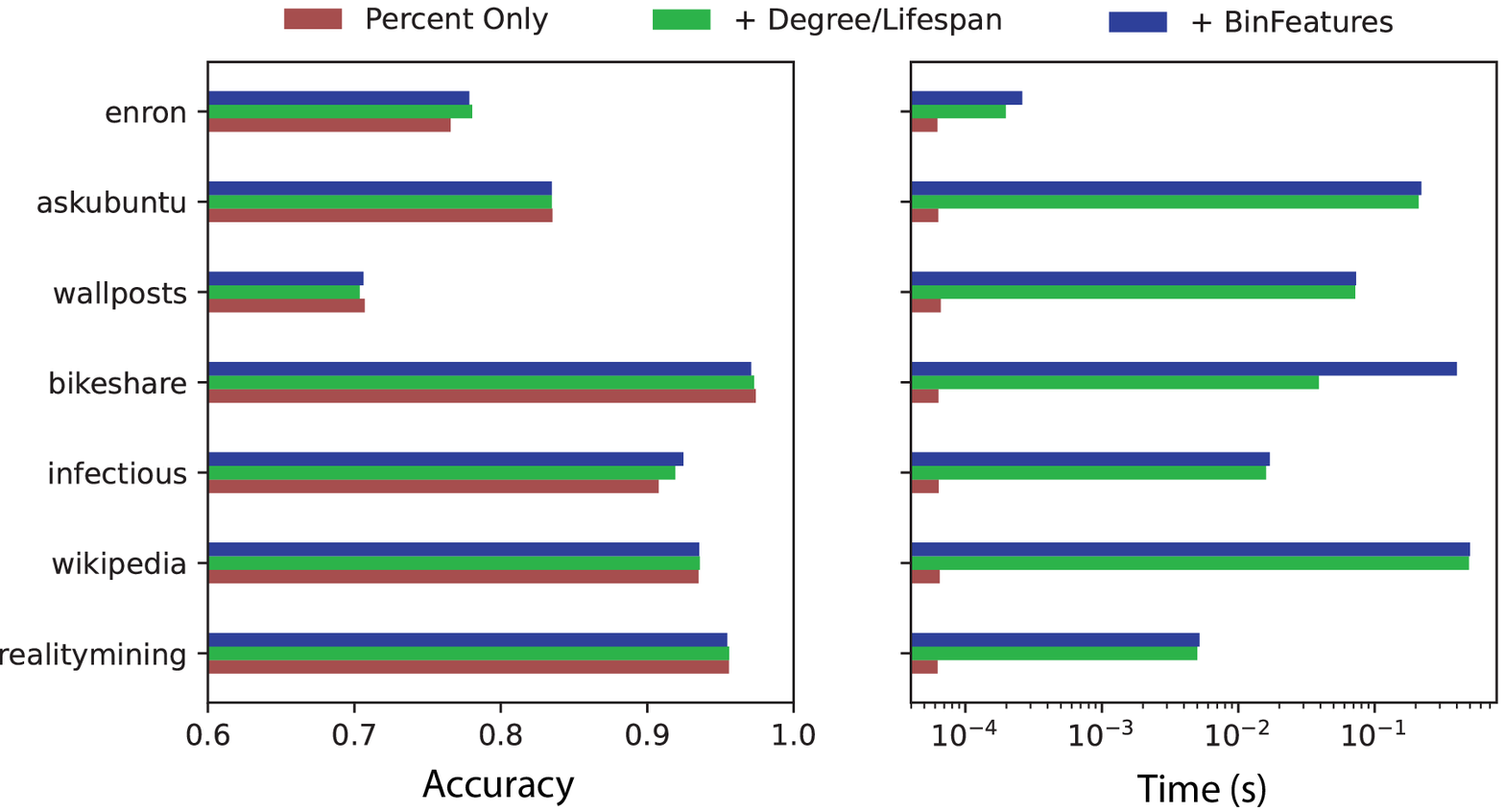}
\caption{Prediction accuracy and feature computation times for the 3 feature sets we consider. 
Using only the node and interval percent features achieves almost the same prediction accuracy as the larger feature sets at a fraction of the computation time.}
\label{fig:feat_times}
\end{figure}

\paragraph{Feature Sets}
We consider the 3 different feature sets we defined in the section \nameref{sec:features}. 
For each feature set, we evaluate both the prediction accuracy of the dual linear regression model trained on 250 slices and the average feature computation time over 5,000 slices. 
The results are shown in Figure \ref{fig:feat_times}. 
Notice that there is hardly any difference in prediction accuracy across the 3 feature sets on all of the data sets except Infectious, where adding more features slightly improves prediction accuracy. 
This suggests that using just the node and interval percent as the features is likely to be sufficient. 

When we examine the computation times of the features, the advantage of using just node and interval percent is much clearer. 
The larger feature sets contain features that take several orders of magnitude more time to compute, as shown in Figure \ref{fig:feat_times}.
This increase in feature computation time would be seen in both the creation time, when generating the training set, and every compound slice prediction thereafter.
Thus, we see no need to use the larger feature sets and continue using only node and interval percent in the rest of this paper.

\begin{figure}[tp]
\centering
\includegraphics[width=4in]{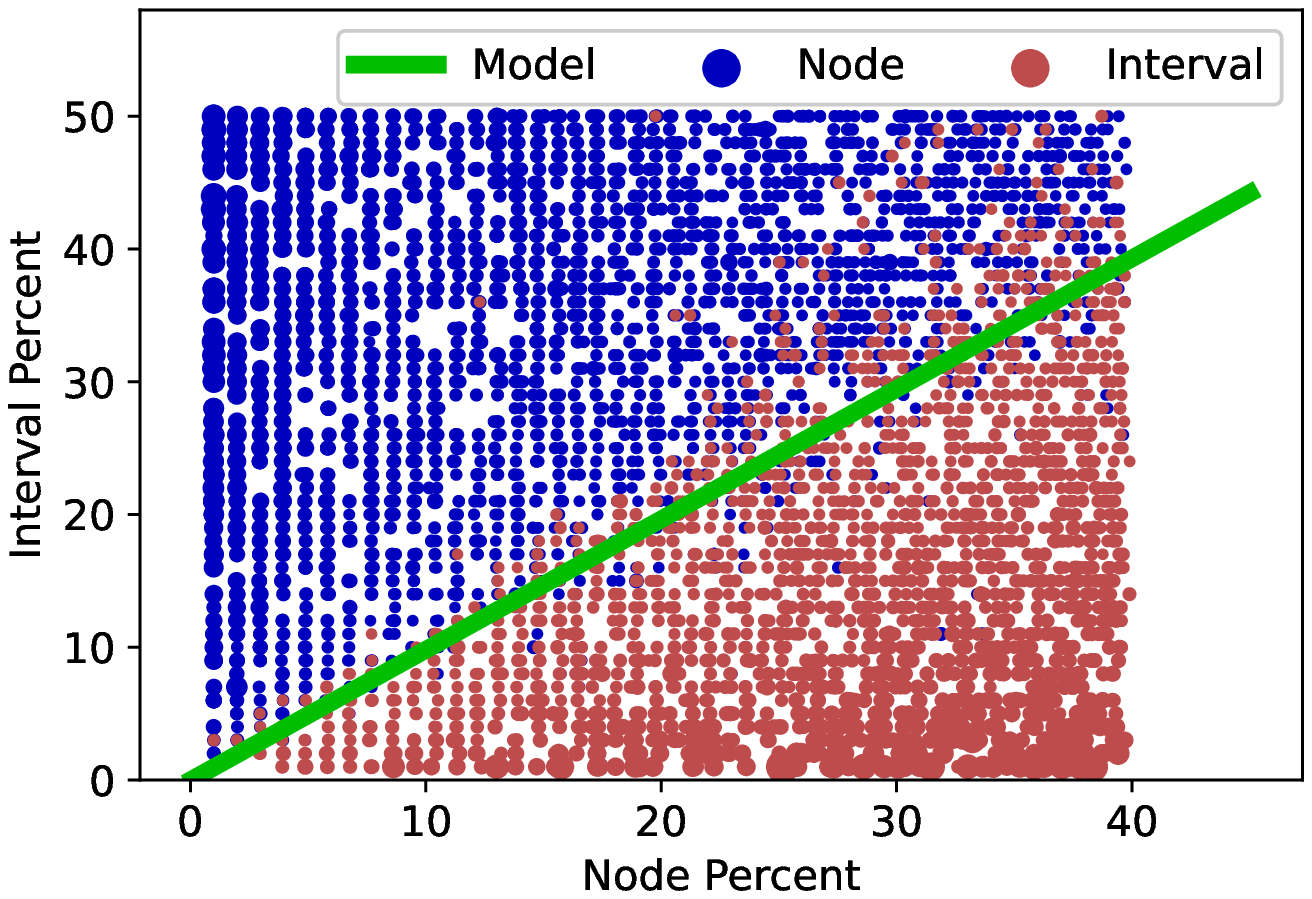}
\caption{Node and interval percent for 5,000 compound slices on the Infectious data set. 
The color of a data point denotes the faster slice, and larger points denote slices where the time difference between node-based and time-based slices is larger.
The green line denotes the threshold for our dual linear regression model predictor: it choose a node-based slice for all points above the line and interval-based slice for all points below the line.}
\label{fig:compound_threshold}
\end{figure}

\paragraph{Exploration of Predictions}
We further explore the predictions made by our predictive model. 
Figure \ref{fig:compound_threshold} shows the two features node percent and interval percent for 5,000 randomly generated compound slices on the Infectious data set. 
The green line denotes the threshold where our dual linear regression model predicts that first performing a node-based or a time-based slice are equally fast. 
For all slices with node and interval percent above the green line, our model predicts that a node-based slice would be faster and vice versa. 
Notice from Figure \ref{fig:compound_threshold} that our model does indeed accurately predict which type of compound slice would be faster in most cases. 
Most of the incorrect predictions occur with slices with small point sizes, which denote a small difference between the two types of compound queries. 
Such incorrect predictions should have a minimal effect on the compound slice times.

\begin{figure}[tp]
\centering
\includegraphics[width=4in]{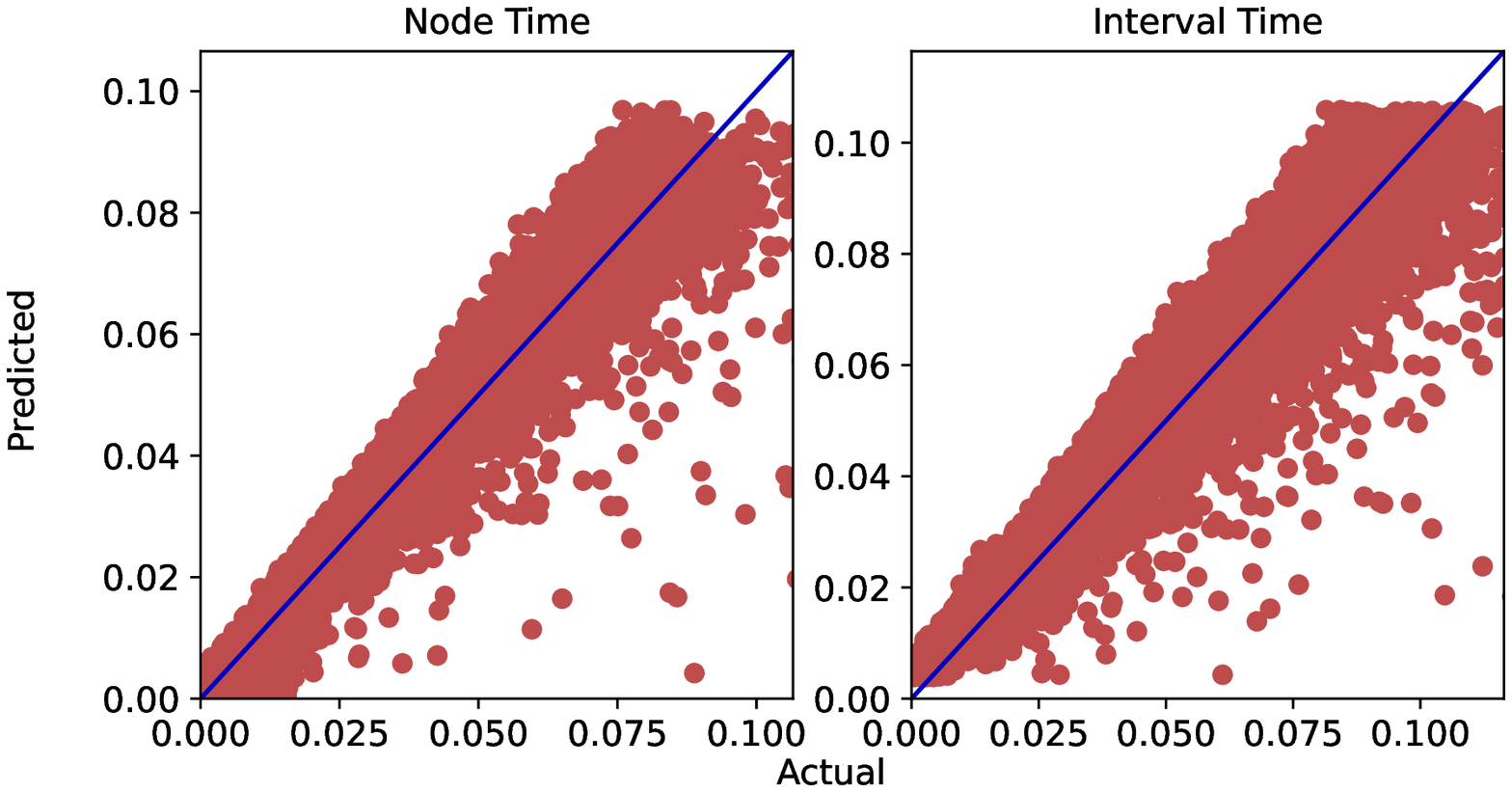}
\caption{Predicted vs.~actual node-based (using adjacency dictionary) and time-based (using interval tree) slice times for 5,000 compound slices on the Infectious data set. 
The blue line illustrates the line $y=x$, denoting perfect predictions. 
Our dual linear regression model predicts both node-based and time-based slice times with reasonable accuracy.}
\label{fig:compound_prediction}
\end{figure}

Figure \ref{fig:compound_prediction} shows the accuracy of our dual linear regression model in predicting the compound slice times first using a node-based slice and first using a time-based slice. 
These predictions are over the same 5,000 compound slices depicted in Figure \ref{fig:compound_threshold}. 
Notice that most of our predictions for both the node-based and time-based slices are fairly accurate, with a small number of outliers where the actual slice time is much larger than predicted. 
These outliers likely denote slices where just the node and interval percent are not sufficient features to accurately predict the compound slice times. 

\subsubsection*{Creation Time}
When adding the predictive compound slicing component to our proposed hybrid data structure, there is an additional component to the creation time---the time required to train the predictor. 
There is also an increase in memory usage, to store the trained predictive model. 

\begin{figure}[tp]
\centering
\includegraphics[width=4in]{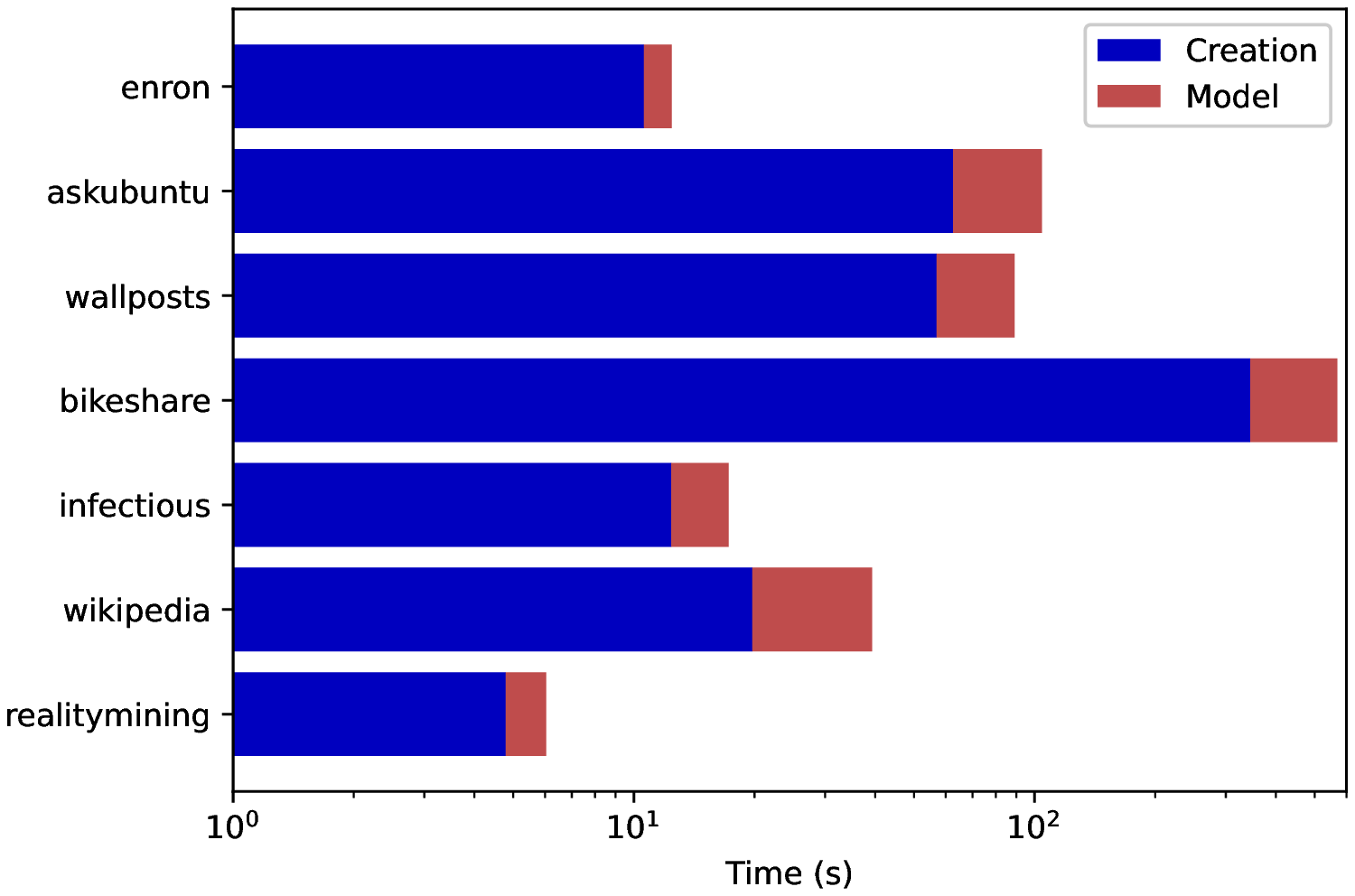}
\caption{Creation time for hybrid data structure with predictive compound slices. 
The time to train the predictive model adds a small overhead to the creation time for the hybrid structure without the predictive component.}
\label{fig:creation_predictive}
\end{figure}

We compute the creation time of IntervalGraph with the predictor on each of the data sets and separate it into the creation time for just the hybrid structure and the time to train the predictive model. 
The results are shown in Figure \ref{fig:creation_predictive}. 
Notice that the added time to train the predictive model does not significantly increase the creation time. 
Actually training the dual linear regression model itself is quite fast; the main component of the added time is the time required to compute the $50$ slices that are used for training. 

\section*{Conclusion}

Temporal networks have the unique capability of capturing the spread of information throughout a network with respect to time.
Analysis of temporal aspects of a network using a dynamic structure can lead to deeper insights that are lost in translation when these networks are flattened into static graphs.
In the interest of increasing our understanding of temporal networks, we propose a hybrid structure that is able to efficiently slice temporal edges using a dimension inaccessible by currently available structures.
Due to its hybrid nature, the proposed structure is still able to benefit from algorithms and techniques developed for static graphs.
The proposed structure achieves a synergistic relationship between its sub-structures by successfully predicting efficient slicing across multiple dimensions.
While these contributions come at the expense of increased creation time and memory usage, the increase is not significant enough to limit viability.
By proposing this new structure, we hope to spark research interests in techniques associated with temporal networks.
We have implemented our proposed hybrid structure in the IntervalGraph class of the DyNetworkX Python package \citep{hilsabeck20} for analyzing dynamic or temporal network data. 



\subsection*{Declarations}


\begin{backmatter}

\section*{Ethics approval and consent to participate}
Not applicable

\section*{Consent for publication}
Not applicable

\section*{Availability of data and materials}
Code and data to reproduce experiments is available at \url{https://github.com/hilsabeckt/hybridtempstruct}.

\section*{Competing interests}
The authors declare that they have no competing interests.

\section*{Funding}
This material is based upon work supported by the National Science Foundation grants IIS-1755824, DMS-1830412, and IIS-2047955. 

\section*{Authors' contributions}
TH developed the predictive compound slicing approach and ran the experiments. 
TH and MA developed the hybrid data structure.
TH and KSX wrote the manuscript and interpreted results. 
All authors read and approved the submitted manuscript.

\section*{Acknowledgements}
None




\bibliographystyle{spbasic} 
\bibliography{references}      

\end{backmatter}
\end{document}